\documentclass[%
 aip,
 amsmath,amssymb,
 reprint,%
]{revtex4-1}

\usepackage{graphicx}
\usepackage{dcolumn}
\usepackage{bm}

\usepackage[utf8]{inputenc}
\usepackage[T1]{fontenc}
\usepackage{mathptmx}
\usepackage{etoolbox}
\usepackage{orcidlink}
\usepackage[parfill]{parskip}
\usepackage{wrapfig}
\usepackage{float}
\usepackage{comment}

\usepackage{float}
\usepackage{txfonts}
\usepackage{gensymb}
\usepackage{adjustbox}
\usepackage{multirow}
\usepackage{epstopdf}
\usepackage{tabulary}
\usepackage{tabularx}
\usepackage{hhline}
\usepackage{booktabs}
\usepackage{array}
\usepackage{subcaption}
\usepackage{titlesec}

\makeatletter
\def\@email#1#2{%
 \endgroup
 \patchcmd{\titleblock@produce}
  {\frontmatter@RRAPformat}
  {\frontmatter@RRAPformat{\produce@RRAP{*#1\href{mailto:#2}{#2}}}\frontmatter@RRAPformat}
  {}{}
}%
\makeatother
\begin{document}

\preprint{AIP/123-QED}

\title[Measuring Regolith Strength in Reduced Gravity Environments in the Laboratory]{Measuring Regolith Strength in Reduced Gravity Environments in the Laboratory}
\author{C. Duffey\,\orcidlink{0009-0000-9901-5269}}
\author{M. Lea}
\affiliation{Florida Space Institute, University of Central Florida, 12354 Research Parkway, Orlando FL-32826, USA.}
\author{J. Brisset}%
\affiliation{Florida Space Institute, University of Central Florida.}

 \email{Christopher.Duffey@ucf.edu}

\date{\today}

\begin{abstract}
This paper presents the design and development of a Shear and Compression Cell (SCC) for measuring the mechanical properties of granular materials in low-gravity environments. This research is motivated by the increasing interest in planetary exploration missions that involve surface interactions, such as those to asteroids and moons. The SCC is designed to measure key mechanical properties, including Young's modulus, angle of internal friction, bulk cohesion, and tensile strength, under both reduced gravity and microgravity conditions. By utilizing a drop tower with interchangeable configurations, we can simulate the gravitational environments of celestial bodies like the Moon and Titan. The SCC, coupled with the drop tower, provides a valuable tool for understanding the behavior of regolith materials and their implications for future space exploration missions. 
\end{abstract}

\maketitle

\section{INTRODUCTION AND BACKGROUND}
\label{s:intro}

Five decades of solar system exploration have taught us that most planetary surfaces are covered in various types of granular materials, so called regolith \citep{housen1982regoliths,mckay1991lunar}. Ever since the first landings on the Moon in the 60s, we have also learned that interacting with this regolith is no easy task. Indeed, astronauts identified lunar dust as a key issue in the safe operation of hardware and manned activities at the surface of the Moon \citep{gaier2007effects}. On other bodies, dust is also a constant challenge for robotic exploration. Not only is dust interfering with nominal operations, like experienced by the InSight rover \citep{lorenz2021first} where covered solar panels were a recurring issue, but the sometimes surprising properties of regolith in these exotic environments also interferes with science activities, such as drilling. The Heat Flow and Physical Properties Package probe (mole) on the InSight rover on Mars required all the ingenuity of the ground operations team to be able to be placed underground: initially foreseen as a drill, the granular properties of the martian soil did not provide necessary friction for the mole to penetrate it, causing the mole to bounce around and form a wide pit around itself rather than dig deeper. Even after implementing a workaround and achieving an underground placement of the instrument, the regolith ended up pushing the mole back out after a few days \citep{spohn2022insight}. Such issues are a clear indication that regolith behaves differently than soil on Earth.

\begin{figure}[b]
    \centering
    \includegraphics[width=0.75\linewidth]{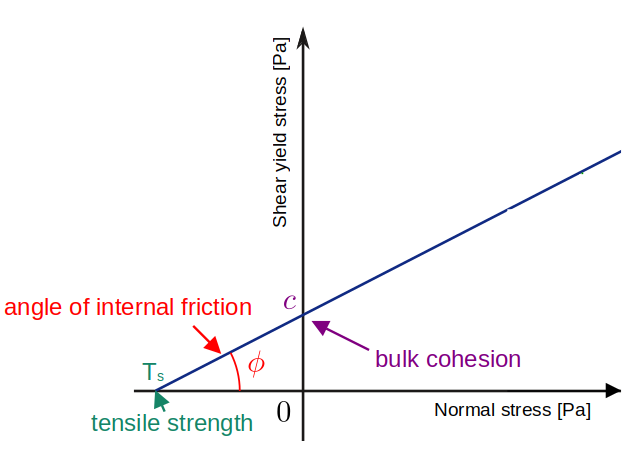}
    \caption{Strength parameters on the Mohr-Coulomb diagram. The shear yield is recorded for varying normal stresses and a linear fit to the data (black line) allows to determine the material's bulk cohesion, angle of internal friction, and tensile strength.}
    \label{f:MCdiagram}
\end{figure}

Now that the Artemis program plans a long-term return of robotic and human activities at the surface of the Moon \citep{bailey2020artemis}, there is an increased interest in understanding the properties of regolith in relevant environments. The scientific community is responding by providing new tools for advancing our understanding of granular material, such as simulants for various types of regoliths \citep{taylor2016evaluations,long2023characterization} as well as ground facilities for testing instrumentation and rover hardware in relevant environments \citep{weber2023developing}. While the intensified lunar exploration is driving a lot of activity, understanding regolith behavior is also of great relevance for the exploration of other solar system bodies, such as Titan with the upcoming Dragonfly mission \citep{barnes2021science}; small rubble pile asteroids with the recent Hayabusa2 \citep{watanabe2017hayabusa2}, OSIRIS-REx \citep{lauretta2017osiris}, and DART missions \citep{rivkin2021double}; and the recently-launched Hera mission \citep{michel2022esa}, which will study how the DART impact affected the Didymos system \citep{lin2023physical}. All of these spacecraft have, or will, interact with the surface regolith of their target, which makes pursuing a better understanding of regolith of key relevance to space exploration.

With this in mind, we have focused some of our efforts on measuring the strength of granular materials in low-gravity environments. We targeted both microgravity for the study of small rubble pile asteroids, such as Ryugu \citep{michikami2019boulder} and Bennu \citep{walsh2019craters}, and partial gravity, for simulating the surfaces of the Moon (0.16$g$, $g$ being the acceleration at the surface of the Earth) and Titan (0.14$g$) among others. We were particularly interested in shear and compression strengths of regolith simulant materials in these extraterrestrial acceleration environments, as this allows for the study of various granular phenomena in the solar system. Examples include the resistance to deformation and disruption in rotating rubble pile asteroids \citep{walsh2018rubble}; the response to impacts and seismic shaking on small bodies and moons \citep{raducan2022reshaping}; the local rheology and thus topography of asteroid and moon surfaces \citep{polishook2020surface}; the challenges and possibilities for construction on the Moon \citep{lim2017extra}; and more.

The Shear and Compression Cell (SCC) design presented in this paper allows for the following measurements:
\begin{itemize}
    \item Young's modulus of the material. By inserting a top plate into the sample, we are measuring its force response. We are thus producing a stress-strain curve that allows for the extraction of the material's Young modulus \citep{briaud2023geotechnical}.
    \item Angle of internal friction (AIF). Using the cell's compression feature, we are able to apply a specific normal stress to the sample. By repeatedly measuring the shear yield of the sample for various normal stresses, we can produce a Mohr-Coulomb diagram and its associated linear fit \citep{mcleod1951mohr}. The slope of this linear fit provides the material's AIF (Figure~\ref{f:MCdiagram}).
    \item Bulk cohesion. In the Mohr-Coulomb diagram, the intersection of the linear fit with the zero normal stress axis provides the bulk cohesion of the material \citep{mcleod1951mohr}.
    \item Tensile strength. In the Mohr-Coulomb diagram, the intersection of the linear fit with the zero shear yield stress axis provides the tensile strength of the material \citep{mcleod1951mohr}.
\end{itemize}

\begin{figure}[t]
    \centering
    \includegraphics[width=0.88\linewidth]{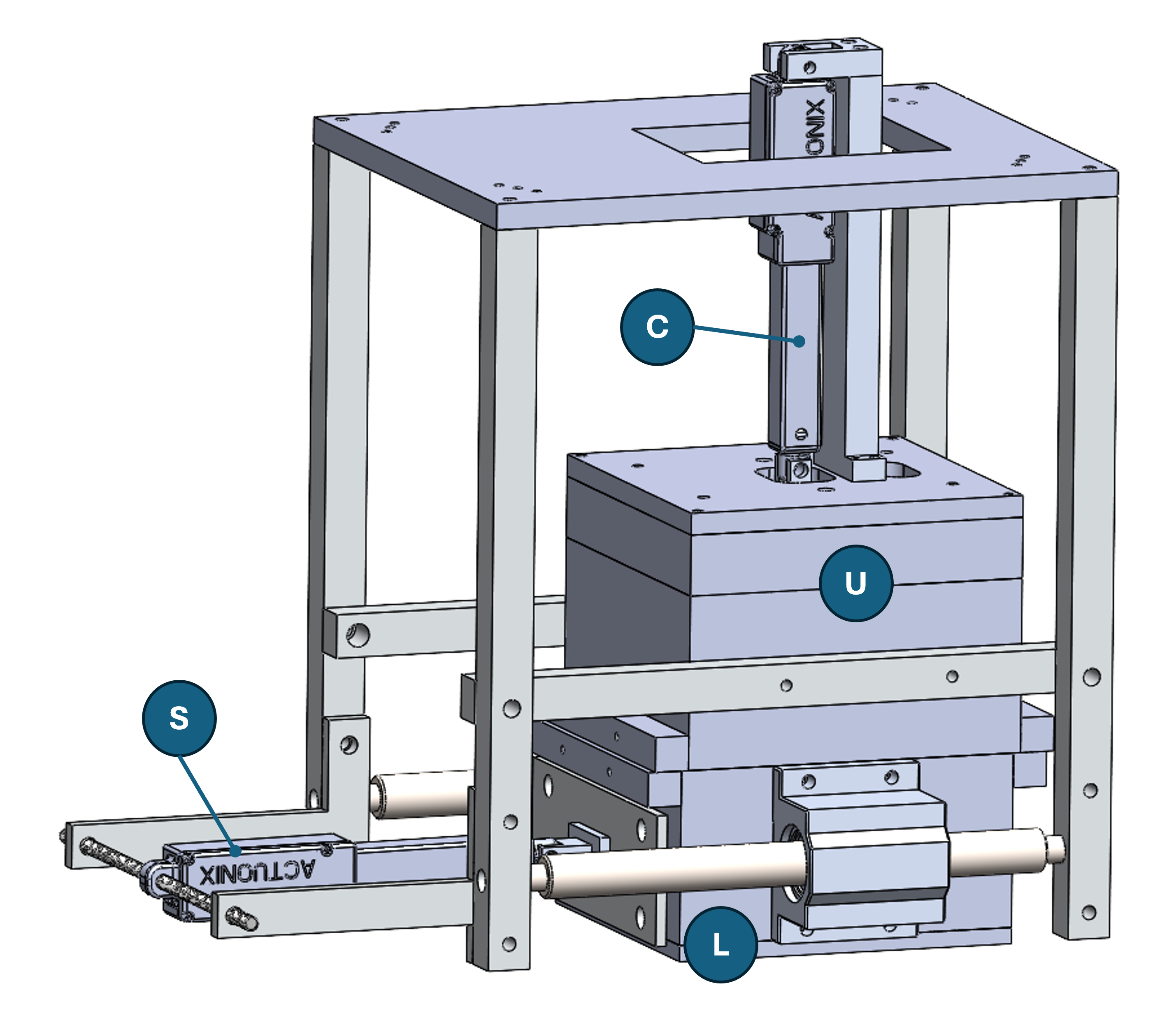}
    \caption{Isometric CAD Model of the SCC unit, showing:  S- Shear Actuator, 
    C- Compression Actuator,  
    U- Upper Shear Box, L- Lower Shear Box, Electronics Unit not shown}
    \label{f:CAD}
\end{figure}

In order to perform our measurements in reduced or microgravity, we built a drop tower with two interchangeable configurations. The tower top plate for the reduced gravity configuration included a breaking motor, which spooled a string attached to the SCC. This motor introduced a chosen acceleration on the dropping SCC, for example 1.35~m/s$^2$ or 0.14$g$ to simulate Titan's surface. For accommodating the SCC to this tower configuration, the cell's top plate was outfitted with a hook where the string was attached. In the microgravity tower configuration, the top plate had an electromagnet  attached to its lower side, which, when turned on, retained a steel plate affixed to the top plate of the SCC. Upon release of the electromagnet, the SCC entered free-fall, thus microgravity. For studying the behavior of the regolith in a gas-free environment, and in particular in order to reduce the influence of air humidity on grain interactions, the microgravity drop tower was designed to be fitted into a vacuum chamber.

In Section~\ref{s:instrument} and \ref{s:towers}, we describe the details of the SCC and the drop tower configurations, respectively. In Section~\ref{s:data}, we provide details on how we calibrate our setup and process data. In Section~\ref{s:measurements}, we describe the measurements we obtained with this setup. Finally, we summarize and conclude in Section~\ref{s:conclusion}.


\begin{figure}[t]
    \centering
    \includegraphics[width=\linewidth]{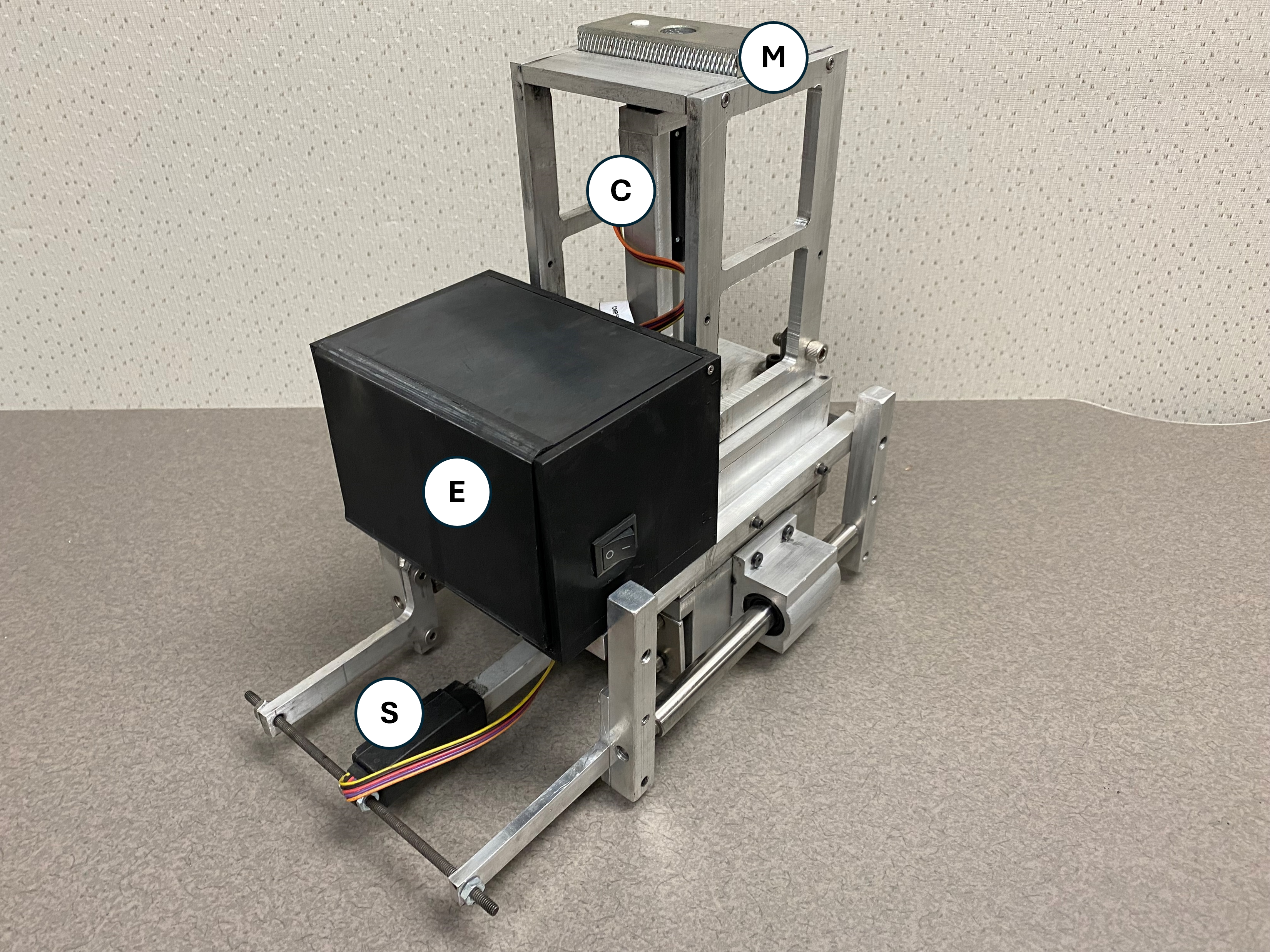}
    \caption{Picture of the completed SCC, showing S- Shear Actuator, 
    C- Compression Actuator,  
    E- Electronics Unit, M- Electromagnet Plate.}
    \label{f:scc_photo}
\end{figure}

\section{INSTRUMENT DESCRIPTION}
\label{s:instrument}

\subsection{Shear and Compression Cell (SCC)}

The SCC Test unit is composed of three major moving parts in a test cell consisting of the upper and lower shear boxes and the compression plate (Figures~\ref{f:CAD}, \ref{f:scc_photo}. A vertically oriented linear actuator is used to compress regolith material loaded in the cell and a horizontally oriented linear actuator moves the lower shear box with respect to the upper shear box, applying a shear force to the loaded regolith material. Two strain gauge type of load cells are installed on each actuator to measure the vertical compression force and the horizontal shear force. The load cells include onboard signal conditioning and temperature compensation, and output their data in engineering units. The load cells are installed in floating mounts that preclude load paths that are not in the shear or compression planes. The shear boxes are connected via a low friction bearing which allows close tolerances between the upper and lower shear boxes to help mitigate leakage of regolith during compression/shear events. The spacing between the upper and lower shear boxes is adjustable.  

The bottom and top of the shear cell are easily removable to load and unload the material under test and to reset the material state following a compression/shear cycle. The test cell holds approximately 800~cm$^3$ of material, however spacers were designed  to allow shearing with less material if desired. The test cell is compatible with several different drop tower configurations which are discussed in detail below.

\begin{figure}[t]
    \centering
    \includegraphics[width=1\linewidth]{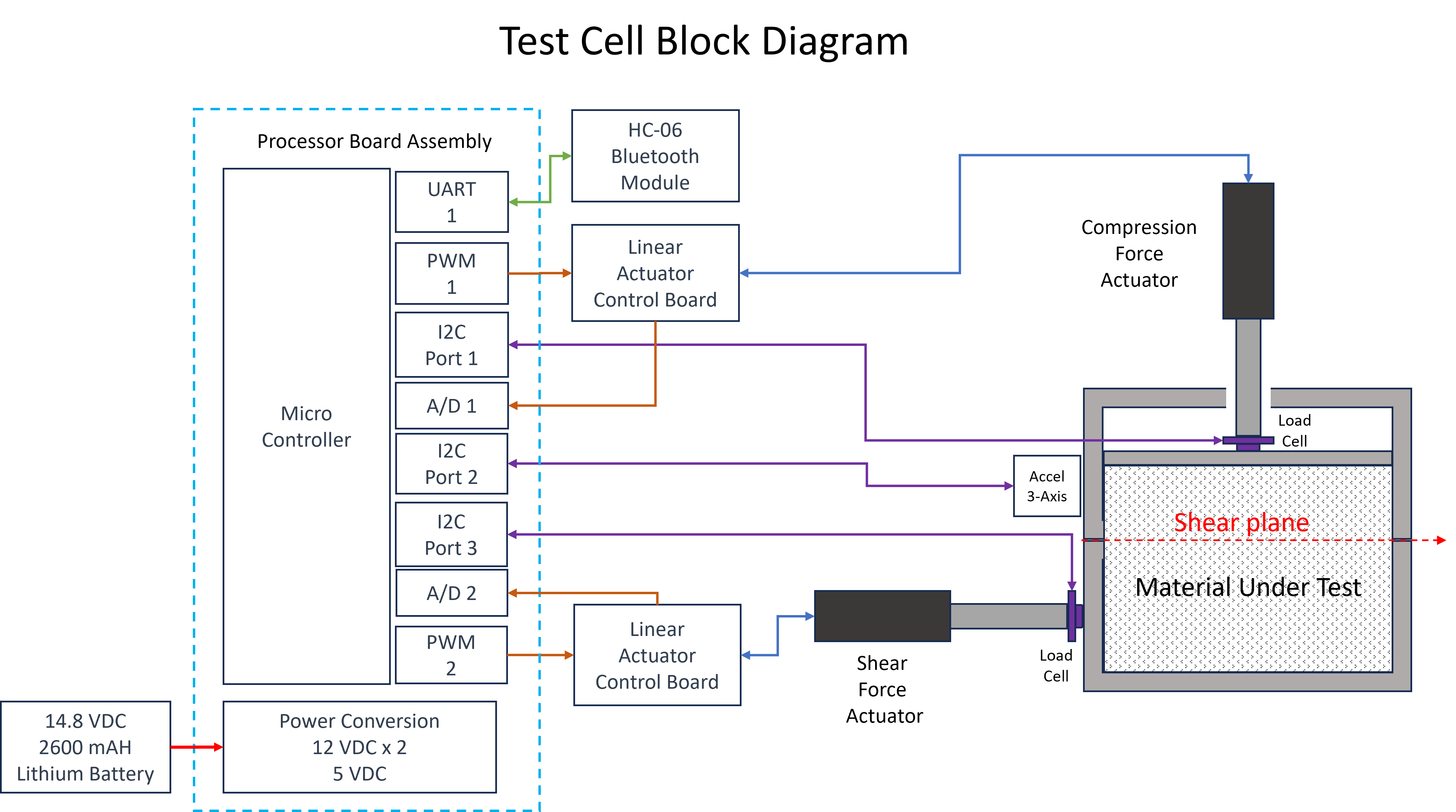}
    \caption{SCC electronics system block diagram.}
    \label{f:electronics}
\end{figure}

\subsection{SCC Controller Electronics}

The test cell controller is based on a high speed microcontroller that orchestrates the various input/output parameters including: the shear and compression actuator commands, the load cells, an inertial measurement unit (IMU), actuator absolute positions and time (Figure~\ref{f:electronics}). The processor formats the data for transmission over a wireless 2.4~GHz Bluetooth link. The microcontroller runs at a 600~MHz clock speed  and is designed to process all of the sensor data: the load cell serial data, analog actuator position A/D channels and the IMU at 50 samples per second, and then stream the data over a wireless link to the data acquisition computer (usually a laptop). A 2600~mAh lithium ion battery can provide up to 20~h of operation using the test cell low power mode. Thanks to the wireless data link and battery power, the test cell is totally self contained and has no external wiring requirements, such as not to introduce any torques on the dropping SCC and preserve the quality of the acceleration environment. The software interface is designed to allow remote calibration and operation and support such features as realignment of the IMU just prior to a drop, reboot of the control processor, entering and leaving low power mode, and independent control of the shear and compression actuators.

The compression and shear load cells utilize a Wheatstone bridge sensor element and integrated processing to calculate the force normal to the sensor in grams - these sensors have integrated calibration and temperature compensation. The data is transmitted out of the sensor on a synchronous serial interface and read by the microcontroller. The actuator control is accomplished in a separate circuit board that implements a typical PID controller, where the PID gains are adjusted to provided a critically damped response with minimal overshoot when the actuator is commanded. A linear, resistive potentiometer outputs an analog voltage that provides feedback to the PID controller as well as an indication of the achieved actuator position and this signal is read by A/D converters and converted to measurements in millimeters.  The IMU provides sensing of XYZ acceleration in units of $g$, and the test cell  orientation pitch, roll, and yaw rotation angles in units of degrees.  A real-time clock can be remotely set from the data acquisition computer and provides highly accurate time tags for the streaming 50 Hz data elements. The test cell has the ability to be put into a lower power "standby" mode to allow for the several hours necessary to pump down the vacuum chamber. When the vacuum chamber has reached its target pressure the test cell is brought out of lower power mode and ready for a drop.

\begin{figure}[t]
    \centering
    \includegraphics[width=1\linewidth]{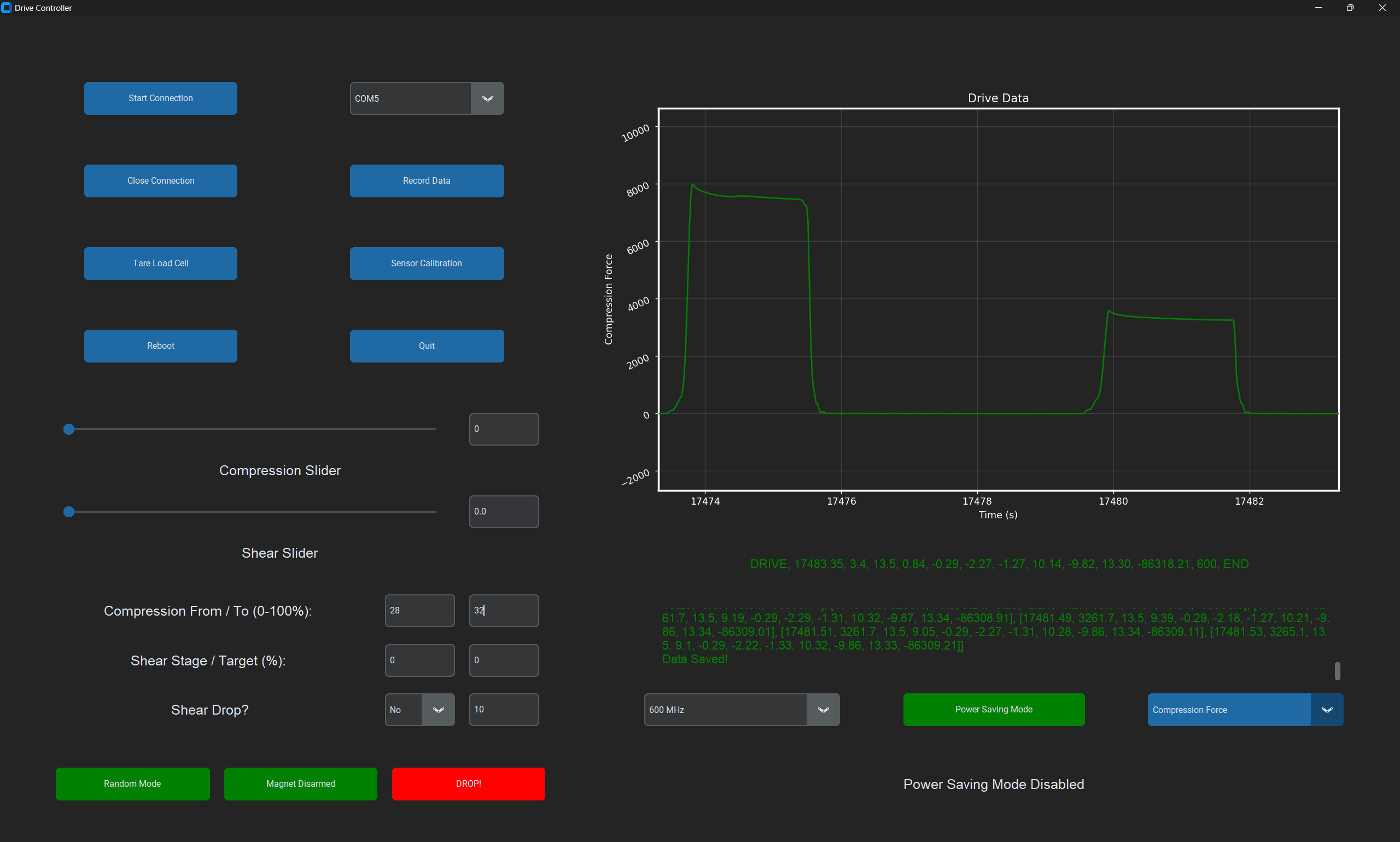}
    \caption{Screen shot of the SCC control interface.}
    \label{f:gui}
\end{figure} 

\subsection{Data Collection}
\label{s:data_collection}
Data Collection is handled by a GUI based controller developed exclusively in Python (Figure~\ref{f:gui}). The controller receives shear cell data via the 2.4~GHz Bluetooth connection, raw data is then parsed and formatted, and primary cell metrics such as compression/shear forces, as well as IMU data are displayed in real-time through a graphical display. A key point for each shear measurement is to ensure the shear yield occurs during reduced gravity/freefall. Therefore there are several adjustable parameters to ensure the correct timing of the measurement. The primary of these parameters is a timing factor within the drop logic loop, which communicates to the controller when to begin shearing. Due to the shorter nature of the microgravity drop tower, this moment timing is slightly before the moment when the electromagnet relay is triggered to release the drop. We have also experimented with utilizing the IMU's accelerometer to sense when a fall is initiated. However, this method does not allow enough time for the sample to have a complete shear within the height of the microgravity tower. For the reduced gravity drop tower, this moment timing can be comfortably after the electromagnetic relay is triggered. In both cases, the instant a drop is triggered, the program splices into the Bluetooth stream and routes it to a separate list. This setup does not require physically retrieving data from the SCC, since all data is wirelessly transferred to the acquisition computer. Once the drop routine is complete, the final calculations are performed and the data is exported into a spreadsheet.

\section{DROP TOWER CONFIGURATIONS}
\label{s:towers}

In the following, we describe the drop tower configurations we designed for micro- and reduced gravity measurements with our SCC. The microgravity setup is outfitted with telescopic legs, extending it from 0.6~m to 0.9~m and allowing it to be installed into a vacuum chamber. The reduced gravity setup is 1.8~m high. It went through a design iteration for providing the desired acceleration environment, first using a counterweight, then a breaking motor.

\subsection{Reduced Gravity Tower}
The first design tested for our reduced gravity tower was based on a classic Atwood machine with counterweight on a pulley calculated  to simulate gravity on the Saturn's moon Titan. An electromagnet was used to  hold the counterweight and electronically release the test cell drop under command of the software. Putting the drop event under software control enables optimization of the shear and compression events timing relative to the desired test $g$-forces.  

The second tower design replaced the counterweight with an adjustable DC motor which enables precise digital control of the drop g-forces. The motor can be adjusted to freewheel during the drop or selectively add a torque to slow the test cell down, this is done by a closed loop using the inertial measurement unit (IMU) to provide a real time measurement of the $g$ forces being experienced by the test cell. The other benefit of this method is that it is invariant to the amount and mass density of the regolith simulant loaded into the test cell. In the Atwood tower design the mass of the samples under test in the cell had to be accounted for when calculating the required counterweight mass needed to achieve the target gravity during the drop.

With a total height of 1.8~m, this reduced gravity drop tower provides 0.48~s at 0.14$g$ (Titan), and 0.49~s at 0.16$g$ (Moon). Since this tower's intent is currently mostly for studies of Titan regolith, it was not designed to fit into our vacuum chamber. Future uses for studies on lunar regolith will require vacuum conditions. This can be achieved by replacing the tower's legs with telescopic ones, as done for the microgravity tower.

\subsection{Microgravity Tower}
\begin{figure}
    \centering
    \includegraphics[width=0.75\linewidth]{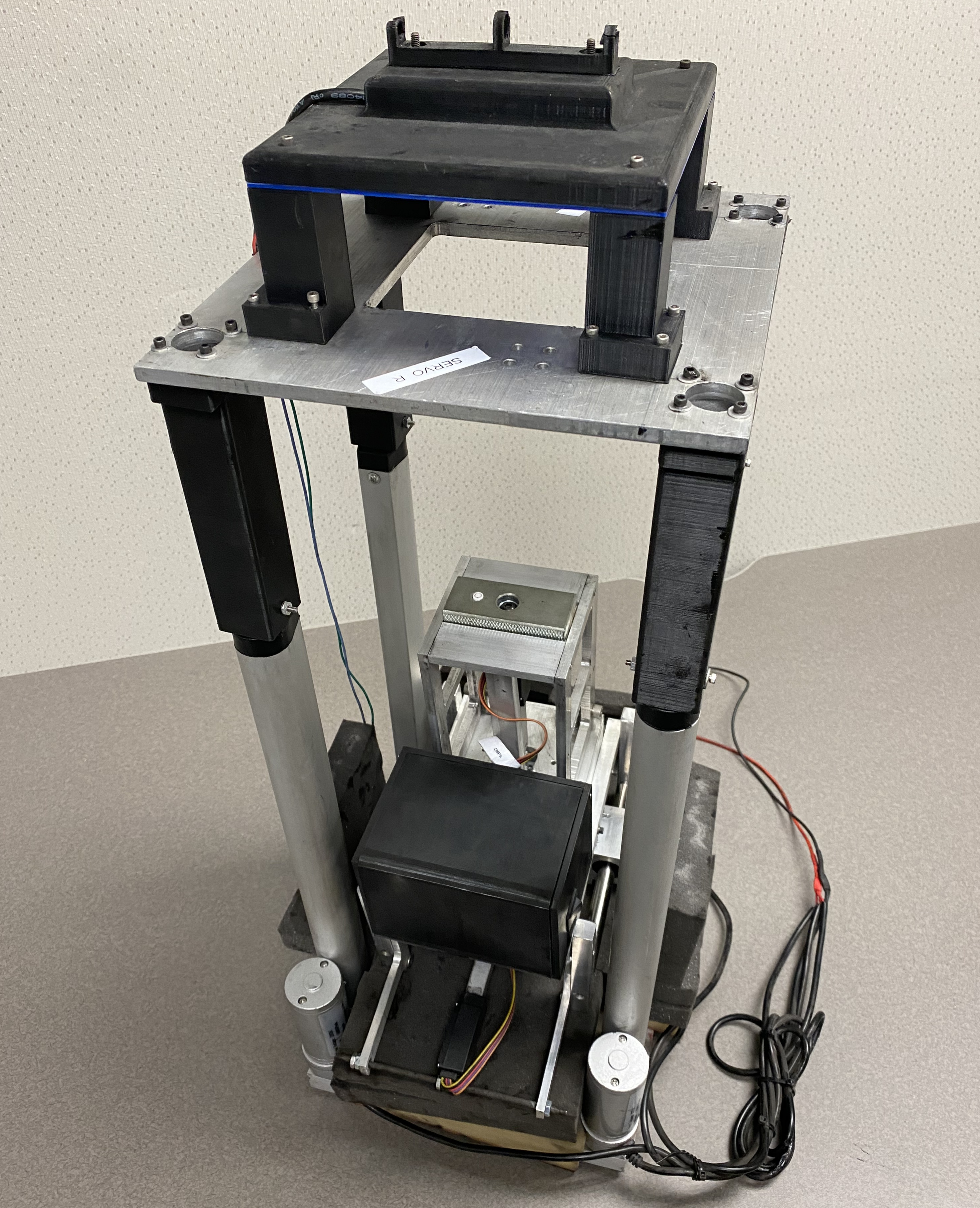}
    \caption{Picture of the SCC at the bottom of the microgravity drop tower outside of the vacuum chamber. There, the telescopic tower legs are retracted all the way.}
    \label{f:drop_tower}
\end{figure}

The microgravity tower employs a more compact design (Figure~\ref{f:drop_tower}). It fits into a vacuum chamber, which allows the elimination of air flow and humidity effects on the simulant behavior. The tower integrates the same electromagnetic release system as the reduced gravity system, but forgoing any gravity mitigation systems. The tower itself consists of four linear actuators which allow for precise control of the towers height. This primarily ensures the tower can both fit within the dimensions of a vacuum chamber, while also maximizing the height and therefore drop distance. An aluminum plate acts as a hub for the actuators and the 3D printed electromagnet housing. The electromagnet is actuated via a micro-controlled relay exterior of the vacuum chamber, wired to its control box via feedthrough flanges.

With a total height of 0.9~m, this tower allows for 0.28~s drops in microgravity.

\section{DATA PROCESSING AND CALIBRATION}
\label{s:data}

\subsection{Data Processing}
\begin{figure}
    \centering
    \includegraphics[width=\linewidth]{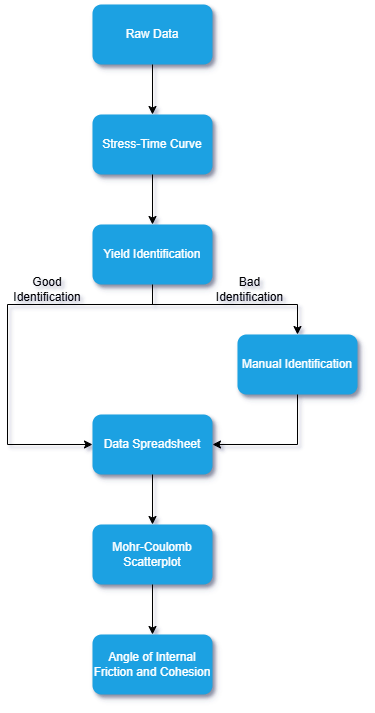}
        \caption{Logical flow diagram of the sample shear yield identification procedure within the data processing software.}
    \label{f:flow_diagram}
\end{figure}

Data Processing is handled both in the back-end of the GUI controller (see Section~\ref{s:data_collection}, and manually via separate python scripts. While these back-end calculations provide the initial shear stress and compression values, additional python scripts are responsible for shear yield identification and data visualization. For shear yield identification, two methods are employed. The first method attempts to use the second derivative of the slope to find the point with the most negative rate of change. This method is completely automated but does not always accurately identify the location of the sample yield on the shear stress-time curve. After each attempt, the shear stress-time curve is shown to the user for verification of the correct identification of the sample yield. In the cases where this method fails to accurately detect the yield, the second method provides a way of manually selecting the yield via user input. On the graph of the shear stress-time profile, the user can interactively select the yield point (Figure~\ref{f:flow_diagram}). The yield points identified by either method are stored into a spreadsheet along with the normal force applied.

This spreadsheet is used to produce a cumulative "Shear Yield vs Normal Force" scatter plot using a separate script. This script then allows for the extraction of the material's angle of internal friction, bulk cohesion, and tensile strength by fitting a linear regression to the yield points.

\subsection{Calibration}

During installation of the SCC onto the drop tower, the load cell calibration is verified by using mass standards to apply a static load to both the compression and shear sensors. If required, a    linear correction can be applied (slope and offset) to refine the the load cell calibration accuracy, and these parameters saved in the controller's non-volatile memory. The actuator travel sensors are also checked using a micrometer to validate the accuracy of the displacement data.  The IMU, that measures the XYZ acceleration and pitch, roll, yaw attitude, automatically calibrates itself to the local gravity vector at SCC power up and can be commanded to re-calibrate by the operator at any time, for instance,  just prior to a drop event.  The IMU utilized in the SCC will exhibit sensor drift over time (typical behavior for a stationary IMU) and calibrating it just prior to a drop will zero out the drift - this is absolutely necessary for vacuum chamber operations where the drop event will happen several hours after power up while the chamber is being pumped down to a vacuum.

\section{MEASUREMENTS OF SIMULATED REGOLITH}
\label{s:measurements}

\subsection{Operational Procedure}

The SCC is designed for simple, reliable, and efficient operation and collection of data. The regolith under test is loaded into the cell with the removal of four fasteners, and while the cell can hold up to 800~cm$^3$ of material, we also manufactured spacers which can be installed as needed to lower the quantity of material required (useful for expensive or hard to obtain simulated regoliths). Once the cell is loaded with material a single switch powers the SCC up and the control computer connects up through the wireless data link. The SCC stores no data internally which eliminates the need to access the electronics for pulling memory cards/flash storage media. Once the cell is loaded with material, the device can be attached to the drop tower and prepared for a drop. If the drop is occurring in a vacuum chamber, the SCC is turned on and put into low power mode during the evacuation of the chamber. Once a drop is commanded and shear/compression data has been collected, the cell needs to be reset for the next measurement. Resetting the cell requires commanding the compression and shear actuators back to their zero starting positions. The cell is then opened using the four top fasteners and the material is mechanically reset. This is done by either stirring it thoroughly or pouring it out and back into the cell in order to break up any residual stresses or compaction induced by the previous measurement. Once this is done, the cell is closed again and ready for another measurement.  

A typical, non-vacuum chamber, reset cycle time is 5 minutes, enabling collection of 12 events in an hour or almost 100 data points in 8 hours. There is some small amount of leakage of material that occurs during a shear measurement, which is related to the material grain size. Very fine materials will exhibit a higher leakage rate than larger grained materials. If a large number of trials is being run, more material can be added during a cell reset if needed.

A measurement in vacuum is limited by the duration of the pumping, typically 1.5~h to reach pressures < 50~mTorr. This allows us to collect approximately 4 data points per day.

\subsection{Shear Measurement}
\begin{figure}[H]
    \centering
    \includegraphics[width=\linewidth,trim={0 0 0 13.5mm},clip]{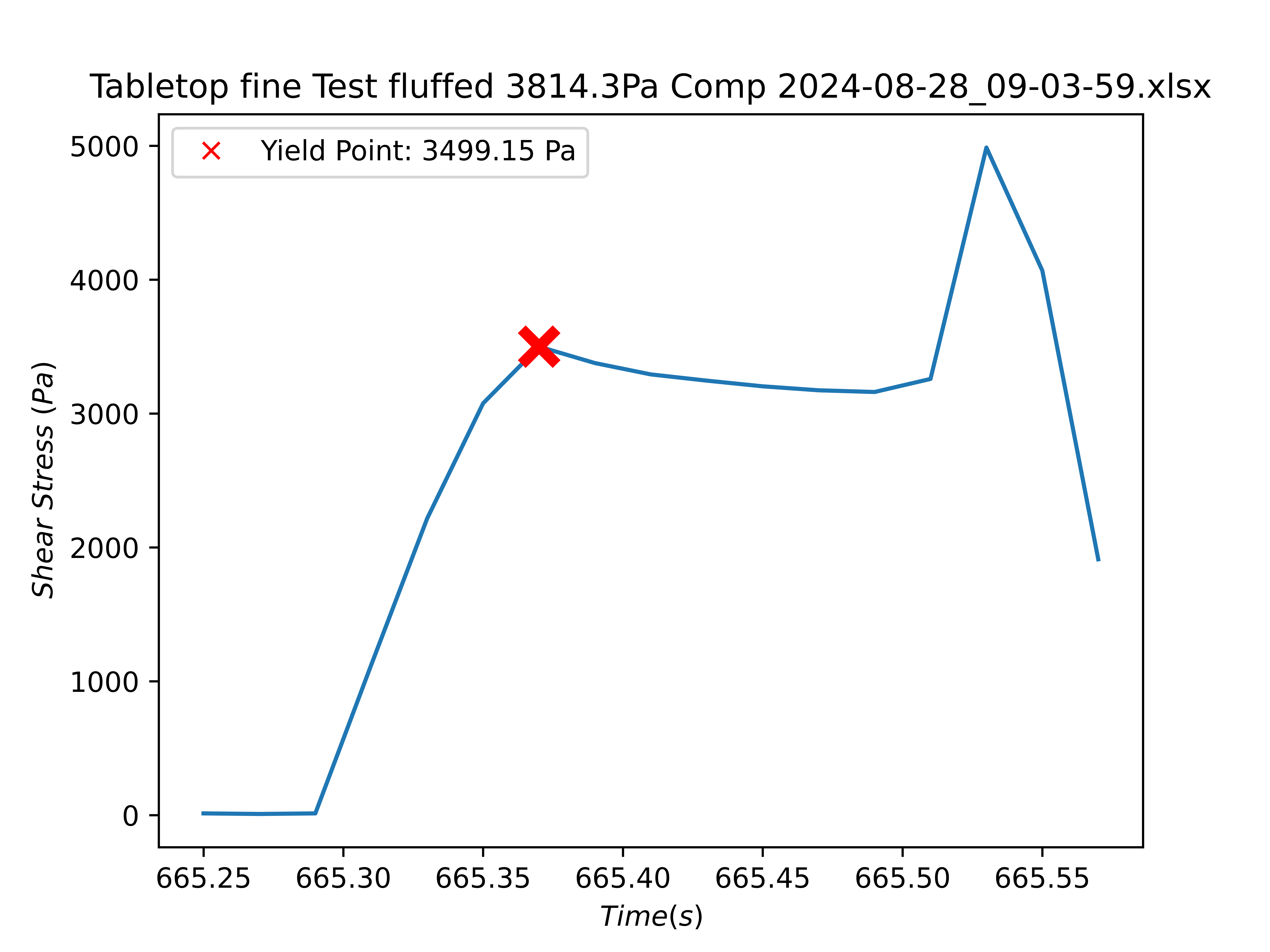}
    \caption{Example of a shear stress-time curve recorded with the SCC in ambient pressure. The sample was composed of Fine grained simulant. Sample yield is marked by a red cross and detected at 3499.15~Pa.}
    \label{f:yield}
\end{figure}

The primary measurements of interest are shear stress over time, which will display the shear yield of the sample (Figure~\ref{f:yield}), as well as the associated compressive strain. These quantities are calculated from the raw force and displacement data upon drop completion. The stress ($\sigma$) is defined as the force per unit area.
\begin{equation}
    \sigma = \frac{Force}{Area}
\end{equation}
The force measurements are provided by the SCC's load cells, with the area being the measured cross-sectional area of the shear cell. For shear measurements, the user enters the desired normal force to be applied to the sample during shearing into the system. The compression actuator then applies the specified normal force to the sample, and after a brief delay to allow the force to normalize, the shear actuator shears the sample. Both actuators then go back to their initial position, and the data is exported.

\subsection{Compression Measurement}
The strain ($\epsilon$) is defined as the amount of deformation ($\delta l$) experienced by the body in the direction of force applied, divided by the initial dimensions of the body ($L$):

\begin{equation}
    \epsilon = \frac{\delta l}{L}
\end{equation}

\begin{figure}[H]
    \centering
    \includegraphics[width=\linewidth]{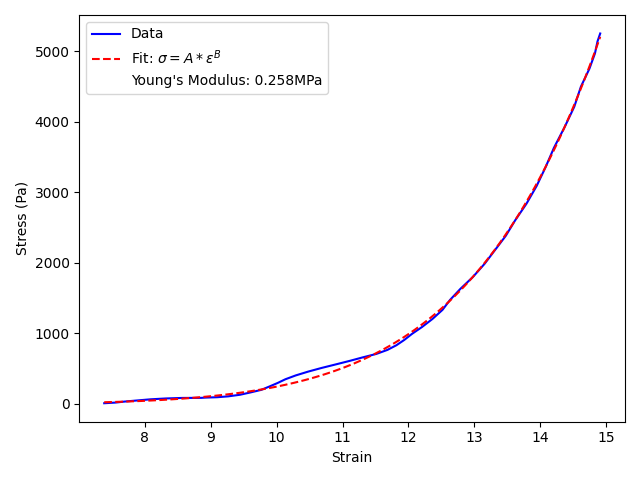}
    \caption{Example of a compression stress-strain curve measured with the SCC in ambient pressure. The sample was composed of Fine Grained simulant. A fit to this curve provides a Young Modulus of 0.258~MPa.}
    \label{f:compression}
\end{figure}

The initial length $L$ is taken as the actuator displacement the the very moment the pressure plate contacts the simulant or where the compression force starts building. The amount of deformation $\delta l$ is taken to be the actuator displacement at peak compression. The stress-strain curve during insertion of the top plate into the sample allows to calculate the Young modulus of the sample material (Figure~\ref{f:compression}). 

For these measurements, a specific normal compression force, or range of forces can be entered into the controller. Data recording begins briefly before the compression actuator initiates, after which, logic within the controller analyzes the instantaneous compression force and increments the actuator position until the specified force is achieved.

\section{SUMMARY AND CONCLUSION}
\label{s:conclusion}

In this paper, we have presented a measurement device that allows us to determine the shear and compression strength of granular material samples at various levels of ambient gravity. This device is composed of a classical shear cell, which has been automated for remotely imposing a specific normal stress and shear rate, and which has been adapted for use in low-$g$ and micro-$g$ drop towers in the laboratory.

We find that this shear and compression cell was able to detect material yield in shear, as well as compression response profiles, during drop tower falls. This will allow us to study granular material strength in acceleration environments relevant to the surfaces of various planetary bodies in the solar system, such as the Moon, Titan, but also small rubble pile asteroids, such as Ryugu and Bennu.

Ideas for future improvements include increasing the speed of the compression and shear actuators to enable using shorter drop towers and further miniaturization of the electronics/battery into an assembly that could be easily removed to facilitate cryogenic cool down of the SCC without potential damage to the lithium battery and stressing the commercial electronic components.

\section{CONFLICT OF INTEREST STATEMENT}
\label{s:coi statement}

The authors have no conflicts to disclose. 

\section{AUTHORS CONTRIBUTIONS}
\label{s:authors contributions}

\begin{itemize}
    \item C.  Duffey\textbf{:} Methodology, Investigation, Software, Formal Analysis, Visualization, Writing - Original Draft, Writing - Review \& Editing
    \item {M. Lea:} Methodology, Investigation, Software, Data Curation, Formal Analysis, Visualization, Writing - Review \& Editing
    \item J. Brisset: Conceptualization, Methodology, Formal Analysis, Supervision, Funding Acquisition, Project Administration, Writing - Review \& Editing
\end{itemize}

  \section{DATA AVAILABILITY STATEMENT}
\label{s:data availability statement}

The data that supports the findings in this paper is available from the authors upon reasonable request

 \section{ACKNOWLEDGMENTS}
\label{s:acknowledgements}

This work acknowledges funding from the NSF AAG grant 1830609 and by the NASA CDAP grant 80NSSC21K0528.

\bibliography{literature}

\begin{thebibliography}{23}%
\makeatletter
\providecommand \@ifxundefined [1]{%
 \@ifx{#1\undefined}
}%
\providecommand \@ifnum [1]{%
 \ifnum #1\expandafter \@firstoftwo
 \else \expandafter \@secondoftwo
 \fi
}%
\providecommand \@ifx [1]{%
 \ifx #1\expandafter \@firstoftwo
 \else \expandafter \@secondoftwo
 \fi
}%
\providecommand \natexlab [1]{#1}%
\providecommand \enquote  [1]{``#1''}%
\providecommand \bibnamefont  [1]{#1}%
\providecommand \bibfnamefont [1]{#1}%
\providecommand \citenamefont [1]{#1}%
\providecommand \href@noop [0]{\@secondoftwo}%
\providecommand \href [0]{\begingroup \@sanitize@url \@href}%
\providecommand \@href[1]{\@@startlink{#1}\@@href}%
\providecommand \@@href[1]{\endgroup#1\@@endlink}%
\providecommand \@sanitize@url [0]{\catcode `\\12\catcode `\$12\catcode `\&12\catcode `\#12\catcode `\^12\catcode `\_12\catcode `\%12\relax}%
\providecommand \@@startlink[1]{}%
\providecommand \@@endlink[0]{}%
\providecommand \url  [0]{\begingroup\@sanitize@url \@url }%
\providecommand \@url [1]{\endgroup\@href {#1}{\urlprefix }}%
\providecommand \urlprefix  [0]{URL }%
\providecommand \Eprint [0]{\href }%
\providecommand \doibase [0]{http://dx.doi.org/}%
\providecommand \selectlanguage [0]{\@gobble}%
\providecommand \bibinfo  [0]{\@secondoftwo}%
\providecommand \bibfield  [0]{\@secondoftwo}%
\providecommand \translation [1]{[#1]}%
\providecommand \BibitemOpen [0]{}%
\providecommand \bibitemStop [0]{}%
\providecommand \bibitemNoStop [0]{.\EOS\space}%
\providecommand \EOS [0]{\spacefactor3000\relax}%
\providecommand \BibitemShut  [1]{\csname bibitem#1\endcsname}%
\let\auto@bib@innerbib\@empty
\bibitem [{\citenamefont {Housen}\ and\ \citenamefont {Wilkening}(1982)}]{housen1982regoliths}%
  \BibitemOpen
  \bibfield  {author} {\bibinfo {author} {\bibfnamefont {K.~R.}\ \bibnamefont {Housen}}\ and\ \bibinfo {author} {\bibfnamefont {L.~L.}\ \bibnamefont {Wilkening}},\ }\bibfield  {title} {\enquote {\bibinfo {title} {Regoliths on small bodies in the solar system},}\ }\href@noop {} {\bibfield  {journal} {\bibinfo  {journal} {In: Annual review of earth and planetary sciences. Volume 10.(A82-35776 17-88) Palo Alto, CA, Annual Reviews, Inc., 1982, p. 355-376.}\ }\textbf {\bibinfo {volume} {10}},\ \bibinfo {pages} {355--376} (\bibinfo {year} {1982})}\BibitemShut {NoStop}%
\bibitem [{\citenamefont {McKay}\ \emph {et~al.}(1991)\citenamefont {McKay}, \citenamefont {Heiken}, \citenamefont {Basu}, \citenamefont {Blanford}, \citenamefont {Simon}, \citenamefont {Reedy}, \citenamefont {French},\ and\ \citenamefont {Papike}}]{mckay1991lunar}%
  \BibitemOpen
  \bibfield  {author} {\bibinfo {author} {\bibfnamefont {D.~S.}\ \bibnamefont {McKay}}, \bibinfo {author} {\bibfnamefont {G.}~\bibnamefont {Heiken}}, \bibinfo {author} {\bibfnamefont {A.}~\bibnamefont {Basu}}, \bibinfo {author} {\bibfnamefont {G.}~\bibnamefont {Blanford}}, \bibinfo {author} {\bibfnamefont {S.}~\bibnamefont {Simon}}, \bibinfo {author} {\bibfnamefont {R.}~\bibnamefont {Reedy}}, \bibinfo {author} {\bibfnamefont {B.~M.}\ \bibnamefont {French}}, \ and\ \bibinfo {author} {\bibfnamefont {J.}~\bibnamefont {Papike}},\ }\bibfield  {title} {\enquote {\bibinfo {title} {The lunar regolith},}\ }\href@noop {} {\bibfield  {journal} {\bibinfo  {journal} {Lunar sourcebook}\ }\textbf {\bibinfo {volume} {567}},\ \bibinfo {pages} {285--356} (\bibinfo {year} {1991})}\BibitemShut {NoStop}%
\bibitem [{\citenamefont {Gaier}(2007)}]{gaier2007effects}%
  \BibitemOpen
  \bibfield  {author} {\bibinfo {author} {\bibfnamefont {J.~R.}\ \bibnamefont {Gaier}},\ }\href@noop {} {\enquote {\bibinfo {title} {The effects of lunar dust on eva systems during the apollo missions},}\ }\bibinfo {type} {Tech. Rep.}\ (\bibinfo {year} {2007})\BibitemShut {NoStop}%
\bibitem [{\citenamefont {Lorenz}, \citenamefont {Lemmon},\ and\ \citenamefont {Maki}(2021)}]{lorenz2021first}%
  \BibitemOpen
  \bibfield  {author} {\bibinfo {author} {\bibfnamefont {R.~D.}\ \bibnamefont {Lorenz}}, \bibinfo {author} {\bibfnamefont {M.~T.}\ \bibnamefont {Lemmon}}, \ and\ \bibinfo {author} {\bibfnamefont {J.}~\bibnamefont {Maki}},\ }\bibfield  {title} {\enquote {\bibinfo {title} {First mars year of observations with the insight solar arrays: Winds, dust devil shadows, and dust accumulation},}\ }\href@noop {} {\bibfield  {journal} {\bibinfo  {journal} {Icarus}\ }\textbf {\bibinfo {volume} {364}},\ \bibinfo {pages} {114468} (\bibinfo {year} {2021})}\BibitemShut {NoStop}%
\bibitem [{\citenamefont {Spohn}\ \emph {et~al.}(2022)\citenamefont {Spohn}, \citenamefont {Hudson}, \citenamefont {Witte}, \citenamefont {Wippermann}, \citenamefont {Wisniewski}, \citenamefont {Kedziora}, \citenamefont {Vrettos}, \citenamefont {Lorenz}, \citenamefont {Golombek}, \citenamefont {Lichtenheldt} \emph {et~al.}}]{spohn2022insight}%
  \BibitemOpen
  \bibfield  {author} {\bibinfo {author} {\bibfnamefont {T.}~\bibnamefont {Spohn}}, \bibinfo {author} {\bibfnamefont {T.~L.}\ \bibnamefont {Hudson}}, \bibinfo {author} {\bibfnamefont {L.}~\bibnamefont {Witte}}, \bibinfo {author} {\bibfnamefont {T.}~\bibnamefont {Wippermann}}, \bibinfo {author} {\bibfnamefont {L.}~\bibnamefont {Wisniewski}}, \bibinfo {author} {\bibfnamefont {B.}~\bibnamefont {Kedziora}}, \bibinfo {author} {\bibfnamefont {C.}~\bibnamefont {Vrettos}}, \bibinfo {author} {\bibfnamefont {R.~D.}\ \bibnamefont {Lorenz}}, \bibinfo {author} {\bibfnamefont {M.}~\bibnamefont {Golombek}}, \bibinfo {author} {\bibfnamefont {R.}~\bibnamefont {Lichtenheldt}},  \emph {et~al.},\ }\bibfield  {title} {\enquote {\bibinfo {title} {The insight-hp3 mole on mars: Lessons learned from attempts to penetrate to depth in the martian soil},}\ }\href@noop {} {\bibfield  {journal} {\bibinfo  {journal} {Advances in Space Research}\ }\textbf {\bibinfo {volume} {69}},\ \bibinfo {pages} {3140--3163} (\bibinfo {year}
  {2022})}\BibitemShut {NoStop}%
\bibitem [{\citenamefont {Bailey}\ \emph {et~al.}(2020)\citenamefont {Bailey}, \citenamefont {Bleacher}, \citenamefont {Lawrence}, \citenamefont {Mahoney},\ and\ \citenamefont {Robinson}}]{bailey2020artemis}%
  \BibitemOpen
  \bibfield  {author} {\bibinfo {author} {\bibfnamefont {B.}~\bibnamefont {Bailey}}, \bibinfo {author} {\bibfnamefont {J.}~\bibnamefont {Bleacher}}, \bibinfo {author} {\bibfnamefont {S.}~\bibnamefont {Lawrence}}, \bibinfo {author} {\bibfnamefont {E.}~\bibnamefont {Mahoney}}, \ and\ \bibinfo {author} {\bibfnamefont {J.~A.}\ \bibnamefont {Robinson}},\ }\bibfield  {title} {\enquote {\bibinfo {title} {The artemis program: Enabling human exploration of the moon},}\ }\href@noop {} {\bibfield  {journal} {\bibinfo  {journal} {The Impact of Lunar Dust on Human Exploration}\ } (\bibinfo {year} {2020})}\BibitemShut {NoStop}%
\bibitem [{\citenamefont {Taylor}, \citenamefont {Pieters},\ and\ \citenamefont {Britt}(2016)}]{taylor2016evaluations}%
  \BibitemOpen
  \bibfield  {author} {\bibinfo {author} {\bibfnamefont {L.~A.}\ \bibnamefont {Taylor}}, \bibinfo {author} {\bibfnamefont {C.~M.}\ \bibnamefont {Pieters}}, \ and\ \bibinfo {author} {\bibfnamefont {D.}~\bibnamefont {Britt}},\ }\bibfield  {title} {\enquote {\bibinfo {title} {Evaluations of lunar regolith simulants},}\ }\href@noop {} {\bibfield  {journal} {\bibinfo  {journal} {Planetary and Space Science}\ }\textbf {\bibinfo {volume} {126}},\ \bibinfo {pages} {1--7} (\bibinfo {year} {2016})}\BibitemShut {NoStop}%
\bibitem [{\citenamefont {Long-Fox}\ and\ \citenamefont {Britt}(2023)}]{long2023characterization}%
  \BibitemOpen
  \bibfield  {author} {\bibinfo {author} {\bibfnamefont {J.~M.}\ \bibnamefont {Long-Fox}}\ and\ \bibinfo {author} {\bibfnamefont {D.~T.}\ \bibnamefont {Britt}},\ }\bibfield  {title} {\enquote {\bibinfo {title} {Characterization of planetary regolith simulants for the research and development of space resource technologies},}\ }\href@noop {} {\bibfield  {journal} {\bibinfo  {journal} {Frontiers in Space Technologies}\ }\textbf {\bibinfo {volume} {4}},\ \bibinfo {pages} {1255535} (\bibinfo {year} {2023})}\BibitemShut {NoStop}%
\bibitem [{\citenamefont {Weber}\ \emph {et~al.}(2023)\citenamefont {Weber}, \citenamefont {Easter}, \citenamefont {Conroy}, \citenamefont {Metke},\ and\ \citenamefont {Britt}}]{weber2023developing}%
  \BibitemOpen
  \bibfield  {author} {\bibinfo {author} {\bibfnamefont {L.}~\bibnamefont {Weber}}, \bibinfo {author} {\bibfnamefont {P.}~\bibnamefont {Easter}}, \bibinfo {author} {\bibfnamefont {M.}~\bibnamefont {Conroy}}, \bibinfo {author} {\bibfnamefont {A.}~\bibnamefont {Metke}}, \ and\ \bibinfo {author} {\bibfnamefont {D.}~\bibnamefont {Britt}},\ }\bibfield  {title} {\enquote {\bibinfo {title} {Developing a large-scale lunar regolith test bin with gravity offload capabilities},}\ }\href@noop {} {\bibfield  {journal} {\bibinfo  {journal} {LPI Contributions}\ }\textbf {\bibinfo {volume} {2806}},\ \bibinfo {pages} {1217} (\bibinfo {year} {2023})}\BibitemShut {NoStop}%
\bibitem [{\citenamefont {Barnes}\ \emph {et~al.}(2021)\citenamefont {Barnes}, \citenamefont {Turtle}, \citenamefont {Trainer}, \citenamefont {Lorenz}, \citenamefont {MacKenzie}, \citenamefont {Brinckerhoff}, \citenamefont {Cable}, \citenamefont {Ernst}, \citenamefont {Freissinet}, \citenamefont {Hand} \emph {et~al.}}]{barnes2021science}%
  \BibitemOpen
  \bibfield  {author} {\bibinfo {author} {\bibfnamefont {J.~W.}\ \bibnamefont {Barnes}}, \bibinfo {author} {\bibfnamefont {E.~P.}\ \bibnamefont {Turtle}}, \bibinfo {author} {\bibfnamefont {M.~G.}\ \bibnamefont {Trainer}}, \bibinfo {author} {\bibfnamefont {R.~D.}\ \bibnamefont {Lorenz}}, \bibinfo {author} {\bibfnamefont {S.~M.}\ \bibnamefont {MacKenzie}}, \bibinfo {author} {\bibfnamefont {W.~B.}\ \bibnamefont {Brinckerhoff}}, \bibinfo {author} {\bibfnamefont {M.~L.}\ \bibnamefont {Cable}}, \bibinfo {author} {\bibfnamefont {C.~M.}\ \bibnamefont {Ernst}}, \bibinfo {author} {\bibfnamefont {C.}~\bibnamefont {Freissinet}}, \bibinfo {author} {\bibfnamefont {K.~P.}\ \bibnamefont {Hand}},  \emph {et~al.},\ }\bibfield  {title} {\enquote {\bibinfo {title} {Science goals and objectives for the dragonfly titan rotorcraft relocatable lander},}\ }\href@noop {} {\bibfield  {journal} {\bibinfo  {journal} {The Planetary Science Journal}\ }\textbf {\bibinfo {volume} {2}},\ \bibinfo {pages} {130} (\bibinfo {year}
  {2021})}\BibitemShut {NoStop}%
\bibitem [{\citenamefont {Watanabe}\ \emph {et~al.}(2017)\citenamefont {Watanabe}, \citenamefont {Tsuda}, \citenamefont {Yoshikawa}, \citenamefont {Tanaka}, \citenamefont {Saiki},\ and\ \citenamefont {Nakazawa}}]{watanabe2017hayabusa2}%
  \BibitemOpen
  \bibfield  {author} {\bibinfo {author} {\bibfnamefont {S.-i.}\ \bibnamefont {Watanabe}}, \bibinfo {author} {\bibfnamefont {Y.}~\bibnamefont {Tsuda}}, \bibinfo {author} {\bibfnamefont {M.}~\bibnamefont {Yoshikawa}}, \bibinfo {author} {\bibfnamefont {S.}~\bibnamefont {Tanaka}}, \bibinfo {author} {\bibfnamefont {T.}~\bibnamefont {Saiki}}, \ and\ \bibinfo {author} {\bibfnamefont {S.}~\bibnamefont {Nakazawa}},\ }\bibfield  {title} {\enquote {\bibinfo {title} {Hayabusa2 mission overview},}\ }\href@noop {} {\bibfield  {journal} {\bibinfo  {journal} {Space Science Reviews}\ }\textbf {\bibinfo {volume} {208}},\ \bibinfo {pages} {3--16} (\bibinfo {year} {2017})}\BibitemShut {NoStop}%
\bibitem [{\citenamefont {Lauretta}\ \emph {et~al.}(2017)\citenamefont {Lauretta}, \citenamefont {Balram-Knutson}, \citenamefont {Beshore}, \citenamefont {Boynton}, \citenamefont {Drouet~d’Aubigny}, \citenamefont {DellaGiustina}, \citenamefont {Enos}, \citenamefont {Golish}, \citenamefont {Hergenrother}, \citenamefont {Howell} \emph {et~al.}}]{lauretta2017osiris}%
  \BibitemOpen
  \bibfield  {author} {\bibinfo {author} {\bibfnamefont {D.}~\bibnamefont {Lauretta}}, \bibinfo {author} {\bibfnamefont {S.}~\bibnamefont {Balram-Knutson}}, \bibinfo {author} {\bibfnamefont {E.}~\bibnamefont {Beshore}}, \bibinfo {author} {\bibfnamefont {W.}~\bibnamefont {Boynton}}, \bibinfo {author} {\bibfnamefont {C.}~\bibnamefont {Drouet~d’Aubigny}}, \bibinfo {author} {\bibfnamefont {D.}~\bibnamefont {DellaGiustina}}, \bibinfo {author} {\bibfnamefont {H.}~\bibnamefont {Enos}}, \bibinfo {author} {\bibfnamefont {D.}~\bibnamefont {Golish}}, \bibinfo {author} {\bibfnamefont {C.}~\bibnamefont {Hergenrother}}, \bibinfo {author} {\bibfnamefont {E.}~\bibnamefont {Howell}},  \emph {et~al.},\ }\bibfield  {title} {\enquote {\bibinfo {title} {Osiris-rex: sample return from asteroid (101955) bennu},}\ }\href@noop {} {\bibfield  {journal} {\bibinfo  {journal} {Space Science Reviews}\ }\textbf {\bibinfo {volume} {212}},\ \bibinfo {pages} {925--984} (\bibinfo {year} {2017})}\BibitemShut {NoStop}%
\bibitem [{\citenamefont {Rivkin}\ \emph {et~al.}(2021)\citenamefont {Rivkin}, \citenamefont {Chabot}, \citenamefont {Stickle}, \citenamefont {Thomas}, \citenamefont {Richardson}, \citenamefont {Barnouin}, \citenamefont {Fahnestock}, \citenamefont {Ernst}, \citenamefont {Cheng}, \citenamefont {Chesley} \emph {et~al.}}]{rivkin2021double}%
  \BibitemOpen
  \bibfield  {author} {\bibinfo {author} {\bibfnamefont {A.~S.}\ \bibnamefont {Rivkin}}, \bibinfo {author} {\bibfnamefont {N.~L.}\ \bibnamefont {Chabot}}, \bibinfo {author} {\bibfnamefont {A.~M.}\ \bibnamefont {Stickle}}, \bibinfo {author} {\bibfnamefont {C.~A.}\ \bibnamefont {Thomas}}, \bibinfo {author} {\bibfnamefont {D.~C.}\ \bibnamefont {Richardson}}, \bibinfo {author} {\bibfnamefont {O.}~\bibnamefont {Barnouin}}, \bibinfo {author} {\bibfnamefont {E.~G.}\ \bibnamefont {Fahnestock}}, \bibinfo {author} {\bibfnamefont {C.~M.}\ \bibnamefont {Ernst}}, \bibinfo {author} {\bibfnamefont {A.~F.}\ \bibnamefont {Cheng}}, \bibinfo {author} {\bibfnamefont {S.}~\bibnamefont {Chesley}},  \emph {et~al.},\ }\bibfield  {title} {\enquote {\bibinfo {title} {The double asteroid redirection test (dart): planetary defense investigations and requirements},}\ }\href@noop {} {\bibfield  {journal} {\bibinfo  {journal} {The Planetary Science Journal}\ }\textbf {\bibinfo {volume} {2}},\ \bibinfo {pages} {173} (\bibinfo {year}
  {2021})}\BibitemShut {NoStop}%
\bibitem [{\citenamefont {Michel}\ \emph {et~al.}(2022)\citenamefont {Michel}, \citenamefont {K{\"u}ppers}, \citenamefont {Bagatin}, \citenamefont {Carry}, \citenamefont {Charnoz}, \citenamefont {De~Leon}, \citenamefont {Fitzsimmons}, \citenamefont {Gordo}, \citenamefont {Green}, \citenamefont {H{\'e}rique} \emph {et~al.}}]{michel2022esa}%
  \BibitemOpen
  \bibfield  {author} {\bibinfo {author} {\bibfnamefont {P.}~\bibnamefont {Michel}}, \bibinfo {author} {\bibfnamefont {M.}~\bibnamefont {K{\"u}ppers}}, \bibinfo {author} {\bibfnamefont {A.~C.}\ \bibnamefont {Bagatin}}, \bibinfo {author} {\bibfnamefont {B.}~\bibnamefont {Carry}}, \bibinfo {author} {\bibfnamefont {S.}~\bibnamefont {Charnoz}}, \bibinfo {author} {\bibfnamefont {J.}~\bibnamefont {De~Leon}}, \bibinfo {author} {\bibfnamefont {A.}~\bibnamefont {Fitzsimmons}}, \bibinfo {author} {\bibfnamefont {P.}~\bibnamefont {Gordo}}, \bibinfo {author} {\bibfnamefont {S.~F.}\ \bibnamefont {Green}}, \bibinfo {author} {\bibfnamefont {A.}~\bibnamefont {H{\'e}rique}},  \emph {et~al.},\ }\bibfield  {title} {\enquote {\bibinfo {title} {The esa hera mission: detailed characterization of the dart impact outcome and of the binary asteroid (65803) didymos},}\ }\href@noop {} {\bibfield  {journal} {\bibinfo  {journal} {The planetary science journal}\ }\textbf {\bibinfo {volume} {3}},\ \bibinfo {pages} {160} (\bibinfo {year}
  {2022})}\BibitemShut {NoStop}%
\bibitem [{\citenamefont {Lin}, \citenamefont {Vincent},\ and\ \citenamefont {Ip}(2023)}]{lin2023physical}%
  \BibitemOpen
  \bibfield  {author} {\bibinfo {author} {\bibfnamefont {Z.-Y.}\ \bibnamefont {Lin}}, \bibinfo {author} {\bibfnamefont {J.-B.}\ \bibnamefont {Vincent}}, \ and\ \bibinfo {author} {\bibfnamefont {W.-H.}\ \bibnamefont {Ip}},\ }\bibfield  {title} {\enquote {\bibinfo {title} {Physical properties of the didymos system before and after the dart impact},}\ }\href@noop {} {\bibfield  {journal} {\bibinfo  {journal} {Astronomy \& Astrophysics}\ }\textbf {\bibinfo {volume} {676}},\ \bibinfo {pages} {A116} (\bibinfo {year} {2023})}\BibitemShut {NoStop}%
\bibitem [{\citenamefont {Michikami}\ \emph {et~al.}(2019)\citenamefont {Michikami}, \citenamefont {Honda}, \citenamefont {Miyamoto}, \citenamefont {Hirabayashi}, \citenamefont {Hagermann}, \citenamefont {Irie}, \citenamefont {Nomura}, \citenamefont {Ernst}, \citenamefont {Kawamura}, \citenamefont {Sugimoto} \emph {et~al.}}]{michikami2019boulder}%
  \BibitemOpen
  \bibfield  {author} {\bibinfo {author} {\bibfnamefont {T.}~\bibnamefont {Michikami}}, \bibinfo {author} {\bibfnamefont {C.}~\bibnamefont {Honda}}, \bibinfo {author} {\bibfnamefont {H.}~\bibnamefont {Miyamoto}}, \bibinfo {author} {\bibfnamefont {M.}~\bibnamefont {Hirabayashi}}, \bibinfo {author} {\bibfnamefont {A.}~\bibnamefont {Hagermann}}, \bibinfo {author} {\bibfnamefont {T.}~\bibnamefont {Irie}}, \bibinfo {author} {\bibfnamefont {K.}~\bibnamefont {Nomura}}, \bibinfo {author} {\bibfnamefont {C.~M.}\ \bibnamefont {Ernst}}, \bibinfo {author} {\bibfnamefont {M.}~\bibnamefont {Kawamura}}, \bibinfo {author} {\bibfnamefont {K.}~\bibnamefont {Sugimoto}},  \emph {et~al.},\ }\bibfield  {title} {\enquote {\bibinfo {title} {Boulder size and shape distributions on asteroid ryugu},}\ }\href@noop {} {\bibfield  {journal} {\bibinfo  {journal} {Icarus}\ }\textbf {\bibinfo {volume} {331}},\ \bibinfo {pages} {179--191} (\bibinfo {year} {2019})}\BibitemShut {NoStop}%
\bibitem [{\citenamefont {Walsh}\ \emph {et~al.}(2019)\citenamefont {Walsh}, \citenamefont {Jawin}, \citenamefont {Ballouz}, \citenamefont {Barnouin}, \citenamefont {Bierhaus}, \citenamefont {Connolly~Jr}, \citenamefont {Molaro}, \citenamefont {McCoy}, \citenamefont {Delbo’}, \citenamefont {Hartzell} \emph {et~al.}}]{walsh2019craters}%
  \BibitemOpen
  \bibfield  {author} {\bibinfo {author} {\bibfnamefont {K.}~\bibnamefont {Walsh}}, \bibinfo {author} {\bibfnamefont {E.}~\bibnamefont {Jawin}}, \bibinfo {author} {\bibfnamefont {R.-L.}\ \bibnamefont {Ballouz}}, \bibinfo {author} {\bibfnamefont {O.}~\bibnamefont {Barnouin}}, \bibinfo {author} {\bibfnamefont {E.}~\bibnamefont {Bierhaus}}, \bibinfo {author} {\bibfnamefont {H.}~\bibnamefont {Connolly~Jr}}, \bibinfo {author} {\bibfnamefont {J.}~\bibnamefont {Molaro}}, \bibinfo {author} {\bibfnamefont {T.~J.}\ \bibnamefont {McCoy}}, \bibinfo {author} {\bibfnamefont {M.}~\bibnamefont {Delbo’}}, \bibinfo {author} {\bibfnamefont {C.}~\bibnamefont {Hartzell}},  \emph {et~al.},\ }\bibfield  {title} {\enquote {\bibinfo {title} {Craters, boulders and regolith of (101955) bennu indicative of an old and dynamic surface},}\ }\href@noop {} {\bibfield  {journal} {\bibinfo  {journal} {Nature Geoscience}\ }\textbf {\bibinfo {volume} {12}},\ \bibinfo {pages} {242--246} (\bibinfo {year} {2019})}\BibitemShut {NoStop}%
\bibitem [{\citenamefont {Walsh}(2018)}]{walsh2018rubble}%
  \BibitemOpen
  \bibfield  {author} {\bibinfo {author} {\bibfnamefont {K.~J.}\ \bibnamefont {Walsh}},\ }\bibfield  {title} {\enquote {\bibinfo {title} {Rubble pile asteroids},}\ }\href@noop {} {\bibfield  {journal} {\bibinfo  {journal} {Annual Review of Astronomy and Astrophysics}\ }\textbf {\bibinfo {volume} {56}},\ \bibinfo {pages} {593--624} (\bibinfo {year} {2018})}\BibitemShut {NoStop}%
\bibitem [{\citenamefont {Raducan}\ \emph {et~al.}(2022)\citenamefont {Raducan}, \citenamefont {Jutzi}, \citenamefont {Zhang}, \citenamefont {Orm{\"o}},\ and\ \citenamefont {Michel}}]{raducan2022reshaping}%
  \BibitemOpen
  \bibfield  {author} {\bibinfo {author} {\bibfnamefont {S.-D.}\ \bibnamefont {Raducan}}, \bibinfo {author} {\bibfnamefont {M.}~\bibnamefont {Jutzi}}, \bibinfo {author} {\bibfnamefont {Y.}~\bibnamefont {Zhang}}, \bibinfo {author} {\bibfnamefont {J.}~\bibnamefont {Orm{\"o}}}, \ and\ \bibinfo {author} {\bibfnamefont {P.}~\bibnamefont {Michel}},\ }\bibfield  {title} {\enquote {\bibinfo {title} {Reshaping and ejection processes on rubble-pile asteroids from impacts},}\ }\href@noop {} {\bibfield  {journal} {\bibinfo  {journal} {Astronomy \& Astrophysics}\ }\textbf {\bibinfo {volume} {665}},\ \bibinfo {pages} {L10} (\bibinfo {year} {2022})}\BibitemShut {NoStop}%
\bibitem [{\citenamefont {Polishook}\ and\ \citenamefont {Aharonson}(2020)}]{polishook2020surface}%
  \BibitemOpen
  \bibfield  {author} {\bibinfo {author} {\bibfnamefont {D.}~\bibnamefont {Polishook}}\ and\ \bibinfo {author} {\bibfnamefont {O.}~\bibnamefont {Aharonson}},\ }\bibfield  {title} {\enquote {\bibinfo {title} {Surface slopes of asteroid pairs as indicators of mechanical properties and cohesion},}\ }\href@noop {} {\bibfield  {journal} {\bibinfo  {journal} {Icarus}\ }\textbf {\bibinfo {volume} {336}},\ \bibinfo {pages} {113415} (\bibinfo {year} {2020})}\BibitemShut {NoStop}%
\bibitem [{\citenamefont {Lim}\ \emph {et~al.}(2017)\citenamefont {Lim}, \citenamefont {Prabhu}, \citenamefont {Anand},\ and\ \citenamefont {Taylor}}]{lim2017extra}%
  \BibitemOpen
  \bibfield  {author} {\bibinfo {author} {\bibfnamefont {S.}~\bibnamefont {Lim}}, \bibinfo {author} {\bibfnamefont {V.~L.}\ \bibnamefont {Prabhu}}, \bibinfo {author} {\bibfnamefont {M.}~\bibnamefont {Anand}}, \ and\ \bibinfo {author} {\bibfnamefont {L.~A.}\ \bibnamefont {Taylor}},\ }\bibfield  {title} {\enquote {\bibinfo {title} {Extra-terrestrial construction processes--advancements, opportunities and challenges},}\ }\href@noop {} {\bibfield  {journal} {\bibinfo  {journal} {Advances in Space Research}\ }\textbf {\bibinfo {volume} {60}},\ \bibinfo {pages} {1413--1429} (\bibinfo {year} {2017})}\BibitemShut {NoStop}%
\bibitem [{\citenamefont {Briaud}(2023)}]{briaud2023geotechnical}%
  \BibitemOpen
  \bibfield  {author} {\bibinfo {author} {\bibfnamefont {J.-L.}\ \bibnamefont {Briaud}},\ }\href@noop {} {\emph {\bibinfo {title} {Geotechnical engineering: unsaturated and saturated soils}}}\ (\bibinfo  {publisher} {John Wiley \& Sons},\ \bibinfo {year} {2023})\BibitemShut {NoStop}%
\bibitem [{\citenamefont {McLeod}(1951)}]{mcleod1951mohr}%
  \BibitemOpen
  \bibfield  {author} {\bibinfo {author} {\bibfnamefont {N.~W.}\ \bibnamefont {McLeod}},\ }\bibfield  {title} {\enquote {\bibinfo {title} {The mohr diagram and coulomb equation},}\ }in\ \href@noop {} {\emph {\bibinfo {booktitle} {Highway Research Board Proceedings}}},\ Vol.~\bibinfo {volume} {30}\ (\bibinfo {year} {1951})\BibitemShut {NoStop}%
\end{thebibliography}%

\end{document}